\documentclass[aps,prb,reprint,twocolumn,superscriptaddress,amsmath,amssymb,10pt]{revtex4-2}
\usepackage{graphicx}
\usepackage{epstopdf}
\usepackage{xcolor}
\usepackage[colorlinks, linkcolor=red,citecolor=blue,urlcolor=blue]{hyperref}
\usepackage[all]{hypcap} 
\begin{document}
\title{Unconventional superconductivity and paramagnetic Meissner response triggered by nonlocal pairing interaction in proximitized heterostructures}

\author{A.~A.~Kopasov}
\affiliation{Institute for Physics of Microstructures, Russian Academy of Sciences, 603950 Nizhny Novgorod, GSP-105, Russia}
\affiliation{Lobachevsky State University of Nizhny Novgorod, 603950 Nizhny Novgorod, Russia}
\author{A.~S.~Mel'nikov}
\affiliation{Institute for Physics of Microstructures, Russian Academy of Sciences, 603950 Nizhny Novgorod, GSP-105, Russia}
\affiliation{Lobachevsky State University of Nizhny Novgorod, 603950 Nizhny Novgorod, Russia}
\affiliation{Moscow Institute of Physics and Technology, Dolgoprudnyi, Moscow Region 141701, Russia}

\date{\today}

\begin{abstract}
Proximity phenomena and induced superconducting correlations in heterostructures are shown to be strongly affected by the 
nonlocal nature of the electronic attraction. The latter can trigger the formation of Cooper pairs consisting of electrons localized in
neighbouring layers even in the absence of direct quasiparticle transfer between the layers. We investigate the manifestations of such nonlocal pairing and resulting unconventional induced superconductivity in an exemplary two-dimensional (2D) electronic system coupled to a conventional superconductor. The interplay between the quasiparticle tunneling and spin-triplet interlayer pairing is shown to generate 
the odd-frequency superconducting correlations in the 2D material which give rise to the paramagnetic contribution to the Meissner response and affect the energy resolved quasiparticle density of states. Experimental evidence for the above nonlocal interface pairing would provide new perspectives in engineering the unconventional superconducting correlations in heterostructures. 
\end{abstract}

\maketitle

\section{Introduction}

For more than half a century the physics of proximity phenomena in various superconducting heterostructures remains an attractive research direction both for experimentalists and theoreticians. The key mechanism underlying the proximity effect is known to arise from the electron transfer between the superconducting and nonsuperconducting material which results in the generation of the induced superconducting correlations in the normal subsystem~\cite{McMillan1968}. The structure of these correlations is determined not only by the order parameter of the primary superconductor but also by the properties of the quasiparticle excitations inside the nonsuperconducting material. As a result, manipulating the electronic spectrum of the latter we get a unique possibility to engineer the induced superconducting state. To tune, e.g., the spin structure of Cooper pairs one can exploit the effect of exchange field in ferromagnetic subsystems~\cite{BuzdinRMP2005,BergeretRMP2005} or the spin-orbit effects arising at the interfaces in heterostructures or in non-centrosymmetric materials~\cite{EdelsteinPRB2003,EschrigRPP2015,LinderNP2015,BobkovaJPCM2022,MelnikobPU2022}. On this way we can get very exotic structure of superconducting correlations providing the possibility to control both the equilibrium and transport effects in superconducting heterostructures. These unconventional superconducting correlations are particularly interesting in the context of recent development of the field of topologically protected quantum computations~\cite{NayakRMP2008,DasSarmaNPJQI2015} and superconducting spintronics~\cite{EschrigRPP2015,LinderNP2015,BobkovaJPCM2022,MelnikobPU2022}.  

Is the above mentioned electron transfer between the subsystems the only mechanism underlying the proximity effect in heterostructures? 
An obvious answer to this question is positive provided we disregard the nonlocal nature of the attraction between the electrons responsible for the superconductivity phenomenon. However, in real systems this attractive interaction mediated, e.g., by phonons   
is not necessary local and can in principle bind the electrons even separated by the interface between the materials. In other words the interface impenetrable for electrons can be still transparent for the phonons. Certainly, different crystal lattice structures of contacting solids and, thus, different elastic properties should result in the reflection of the elastic waves incident on the interface. This reflection as well as the screening effects are expected to weaken any attractive forces between the quasiparticles localized in neighbouring subsystems. Still, if this nonlocal attraction is nonzero it can cause the formation of Cooper pairs of electrons positioned, e.g., in neighboring layers of the multilayered structure. This scenario of interlayer pairing is not completely new, of course, and previously  
it was discussed in the context of different layered superconductors such as transition metal dichalcogenides and high-$T_c$ cuprates~\cite{EfetovJETP1975, TesanovicPRB1987, BulaevskiiPRB1990, KlemmLiu, KettemannPRB1992}. An important property of such interlayer pairing is that due to the nonlocality of the Cooper pair wavefunction (or more rigorously, the anomalous Green function) the Pauli principle does no more impose well known severe restrictions on the spins of electrons in the pair~\cite{BCS} which usually hamper the formation of triplet superconducting correlations. Exactly this argument in favour of possible triplet interlayer pairing motivated A.~I.~Larkin and K.~B.~Efetov~\cite{EfetovJETP1975} to consider this type of correlations to explain the extremely high upper critical fields in TaS$_2$ (pyridine) which were shown to exceed the paramagnetic limit~\cite{MorrisPRB1973}. These theoretical considerations of the interlayer pairing have been
further developed~\cite{TesanovicPRB1987, BulaevskiiPRB1990, KlemmLiu, KettemannPRB1992} in the context of extensive studies of superconductivity in cuprates which also can be well described by the model of identical superconducting layers. 

All the above theoretical works were devoted to the study of natural layered compounds and, thus, assumed the coinciding electronic structure of the individual layers.   
The goal of our work is to apply the idea of Larkin and Efetov to the artificial heterostructures where the neighboring layers can possess quite different individual electronic characteristics including the difference in the normal state band spectra as well as different pairing properties. Considering the formation of the pairs consisting of electrons with different band spectra one can immediately notice the formal analogy of this problem to the one describing a standard singlet superconductor with the quasiparticle spectrum split by the Zeeman or exchange field. Certainly, the effective exchange field in our scenario will depend on the quasiparticle momentum but the basic features of the system including the depairing effect of the difference in the electronic spectra, formation of the odd - frequency superconducting correlations and the inhomogeneous Fulde - Ferrell - Larkin - Ovchinnikov (FFLO) state should be similar to the well known models describing the superconductors in the presence of the spin splitting field~\cite{BergeretRMP2005, ChandrasekharAPL1962,ClogstonPRL1962,SaintJames,Fulde1964,LarkinJETP1965} (see also Ref.~\cite{LinderRMP2019} and references therein). Let us emphasize that all these features are expected to appear in heterostructures without any ferromagnetic layers which could provide the source of the true exchange field determined by the interaction of electron spins with ferromagnetic ordering. This observation looks particularly interesting if we remind some rather old experiments indicating the presence of low temperature paramagnetic contribution to the Meissner response in superconducting cylinders covered by thin normal metal layers~\cite{Mota}. Several theoretical works argue that this phenomenon can be associated with the orbital effects~\cite{BruderPRL1998}, the electronic repulsion in the normal metal layer~\cite{FaucherePRL1999}, the appearance of the $p$-wave superconductivity at low temperatures~\cite{MakiPLA2000}, and the effects of the spin-orbit interaction~\cite{EspedalPRL2016}. In view of the above discussion this paramagnetic response could originate also from the odd-frequency superconducting correlations generated by the nonlocal electron pairing according to the Larkin - Efetov mechanism. Another interesting application of the interlayer pairing arises if we consider its role in Majorana - type systems~\cite{AlbrechtN2016,BommerPRL2019} where this mechanism can probably help to get rid of necessity of rather high magnetic fields providing the Zeeman splitting of energy band in Majorana nanowires. Motivated by all these arguments we studied the manifestation of the Larkin - Efetov mechanism in two exemplary systems: (i) a bilayer consisting of thin films with a certain energy shift of the conduction bands; (ii) a two dimensional electron gas (2DEG) placed in contact with a thick superconducting layer (SC).      
   
\textit{Two layer model}.--- 
We proceed with the consideration of the phenomenon of interlayer pairing in a two layer model which can be viewed as the generalization of the one studied previously in~\cite{EfetovJETP1975}. The key point is that we assume the normal quasiparticle spectra to differ by a certain constant shift due to different conduction band offsets. Note that for simplicity we neglect here the Cooper pairing in each individual layer. The total Hamiltonian accounting for the interlayer pairing takes the form: $H = \sum_{j = 1,2}H_j + H_t + H_{\rm int}$, where
\begin{equation}\label{normal_state_Hamiltonian_two_layers}
 H_j = \int d^2\mathbf{r} \ \psi^{\dagger}_{j\sigma}(\mathbf{r})\hat{\xi}_j\psi_{j\sigma}(\mathbf{r}) \ ,
\end{equation}
describe isolated two-dimensional layers, $\sigma = \uparrow,\downarrow$ denotes spin degrees of freedom (summation over repeated indices is implied), $\hat{\xi}_j = -\nabla_{\mathbf{r}}^2/2m - \mu_j$, and $m$ is the effective mass. The relative shift of the conduction bands is expressed as $(\mu_1 - \mu_2) = 2\chi$, where $\mu_j$ is the difference between the chemical potential and the bottom of the corresponding energy band. The tunnel Hamiltonian has the form
\begin{equation}\label{tunneling_Hamiltonian_two_layers}
 H_t = \int d^2\mathbf{r}\left[t\psi_{1\sigma}^{\dagger}(\mathbf{r})\psi_{2\sigma}(\mathbf{r}) + t^*\psi_{2\sigma}^{\dagger}(\mathbf{r})\psi_{1\sigma}(\mathbf{r})\right] \ ,
\end{equation}
and the interlayer electron-electron interaction is described by the term
\begin{equation}\label{interlayer_interaction_Hamiltonian}
 H_{\rm int} = \frac{U_0}{2}\int d^2\mathbf{r} \ \psi_{1\sigma}^{\dagger}(\mathbf{r})\psi_{2\sigma'}^{\dagger}(\mathbf{r})\psi_{2\sigma'}(\mathbf{r})\psi_{1\sigma}(\mathbf{r}) \ .
\end{equation}
Assuming the in-plane translational symmetry and spatially homogeneous interlayer pairing state, we obtain the following system of Gor'kov equations written in the Matsubara frequency - momentum representation~\cite{summplemental}
\begin{equation}\label{Gorkov_eqs_two_layers}
 \begin{bmatrix}-i\omega_n + \check{\tau}_z\xi_{1\mathbf{k}}&\check{t}\\ \check{t}^{\dagger}&-i\omega_n + \check{\tau}_z\xi_{2\mathbf{k}}\end{bmatrix}\begin{bmatrix}\check{G}_{11}&\check{G}_{12}\\ \check{G}_{21}&\check{G}_{22}\end{bmatrix} = 1 \ ,
\end{equation}
where $\omega_n = 2\pi T(n+1/2)$, $T$ is temperature, $n$ is an integer, $\xi_{j\mathbf{k}} = \mathbf{k}^2/2m - \mu_j$, and $\check{\tau}_i$ ($i = x,y,z$) are the Pauli matrices acting in the electron-hole space. The coupling matrix $\check{t}^{\dagger}$, the Green functions of the subsystems ($\check{G}_{11}$ and $\check{G}_{22}$) and the mixed ones ($\check{G}_{12}$ and $\check{G}_{21}$) acquire a nontrivial structure in the particle-hole space
\begin{equation}
 \check{t}^{\dagger} = \begin{bmatrix}t^*&\hat{\Delta}_{\rm int}\\ -\hat{\Delta}_{\rm int}^*&-t\end{bmatrix} \ , \ \ \check{G}_{ij} = \begin{pmatrix}\hat{G}_{ij}&\hat{F}_{ij}\\ \hat{F}^{\dagger}_{ij}&\hat{\bar{G}}_{ij}\end{pmatrix} \ ,
\end{equation}
due to the presence of the interlayer gap function $\hat{\Delta}_{\rm int}$. 

\begin{figure}[htpb]
\centering
\includegraphics[scale = 0.48]{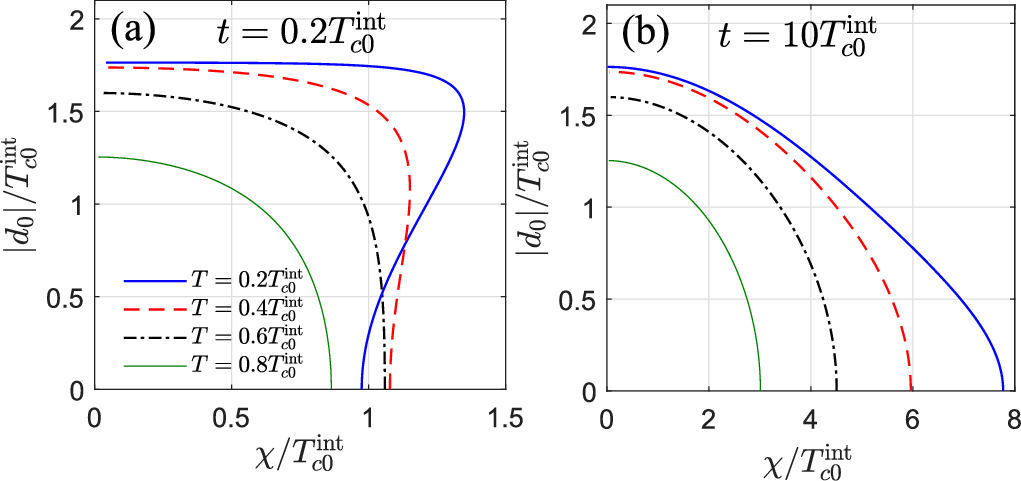}
\caption{The absolute value of the spin-singlet interlayer gap function $|d_0|$ for the two-layer model~(\ref{Gorkov_eqs_two_layers}) vs the band splitting $\chi$ for $T/T_{c0}^{\rm int} = 0.2$, 0.4, 0.6, and 0.8. (a) and (b) correspond to $t/T_{c0}^{\rm int} = 0.2$ and 10, respectively. Here $T_{c0}^{\rm int}$ is the critical temperature for the interlayer order parameter at $\chi = 0$ and $t$ is the tunneling amplitude.}
\label{Fig:self_consistency}
\end{figure} 

We demonstrate the analogy between the effects of the band structure on nonlocal Cooper pairs and the ones of the spin-splitting field in a conventional superconductor by solving the self-consistency equation
\begin{equation}\label{self_consistency_two_layers}
 \hat{\Delta}_{\rm int} = -\frac{U_0}{2}T\sum_{\omega_n}\int\frac{d^2\mathbf{k}}{(2\pi)^2}\hat{F}_{12}(\mathbf{k};\omega_n) \ ,
\end{equation}
for a particular case of the spin-singlet interlayer pairing $\hat{\Delta}_{\rm int} = d_0(i\hat{\sigma}_y)$ and $U_0 = -|U_0|$. 
Substituting the solution of Eq.~(\ref{Gorkov_eqs_two_layers}) into~(\ref{self_consistency_two_layers}), we derive~\cite{summplemental}
\begin{eqnarray}\label{gap_equation_two_layers}
 1 = -\frac{U_0}{4}T\sum_{\omega_n}\int\frac{d^2\mathbf{k}}{(2\pi)^2}\biggl[\frac{1}{\omega_n^2 + E_-^2} + \frac{1}{\omega_n^2 + E_+^2}- \\
 \nonumber
 \frac{\chi^2}{\sqrt{\xi_{\mathbf{k}}^2|t|^2 + \chi^2(\xi_{\mathbf{k}}^2 + |d_0|^2)}}\left(\frac{1}{\omega_n^2 + E_-^2} - \frac{1}{\omega_n^2 + E_+^2}\right)\biggl] \ ,
\end{eqnarray}
where $E_{\pm}(\mathbf{k})$ is the quasiparticle energy spectrum of the two-layer system 
\begin{eqnarray}
 E_{\pm}^2(\mathbf{k}) = |t|^2 + |d_0|^2 + \xi_{\mathbf{k}}^2 + \chi^2 \\
 \nonumber
 \pm 2\sqrt{\xi_{\mathbf{k}}^2(|t|^2 + \chi^2) + \chi^2|d_0|^2} \ ,
\end{eqnarray}
and $\xi_{\mathbf{k}} = (\xi_{1\mathbf{k}} + \xi_{2\mathbf{k}})/2$. The form of the gap equation~(\ref{gap_equation_two_layers}) is similar to the one for the superconductor with Rashba spin-orbit coupling under the influence of the Zeeman field (see, e.g., Eq.~(27) in Ref.~\cite{TewariNJP2011}). Thus, we anticipate that a relative shift of the conduction bands should provide a depairing effect for interlayer Cooper pairs whereas the tunnel coupling mixes the states of isolated layers and should play a role similar to the spin-orbit interaction.

For the solution of the self-consistency equation we assume $\mu_j$ to be much larger than the cut-off energy $\Omega$ and then eliminate the cut-off in favor of the superconducting critical temperature of the interlayer order parameter $T_{c0}^{\rm int}$ at zero conduction band shift $\chi  =0$~\cite{KopninBook}. The resulting gap equation reads 
\begin{eqnarray}
 \ln\left(\frac{T}{T_{c0}^{\rm int}}\right) + 2\pi T {\rm Re} \sum_{\omega_n >0}\biggl[\frac{1}{\omega_n} - \\
 \nonumber
 - \frac{(|t|^2 + i\zeta)}{\zeta\sqrt{-\omega_n^2 - |d_0|^2 + \chi^2 + |t|^2+2i\zeta}}\biggl] = 0 \ ,
\end{eqnarray}
where $\zeta = \sqrt{\omega_n^2(|t|^2 + \chi^2) + |t|^2|d_0|^2}$. Typical $|d_0(\chi)|$ plots for different $T$ shown in Fig.~\ref{Fig:self_consistency} demonstrate the suppression of the interlayer gap function by the band splitting. Fig.~\ref{Fig:self_consistency}(a) shows that for rather low temperatures and weak tunnel couplings there appear $\chi$-regions with more than one solution of the gap equation, which is typical for the paramagnetic effect in superconductors. Thus, by the analogy with the spin-split superconductors~\cite{SaintJames,Fulde1964,LarkinJETP1965} we argue that the relative band shift can lead to the appearance of the odd-frequency interlayer superconducting correlations and the FFLO instability. Fig.~\ref{Fig:self_consistency}(b) shows that the quasiparticle tunnelling suppresses the depairing effect of the band splitting.
Note that if we now consider the joint effect of the relative band shift and the true Zeeman field, one can naturally expect the emergence of the reentrant superconductivity similar to the situation considered in Ref.~\cite{BuzdinPRL2005}.

\textit{2DEG in contact with a thick $s$-wave superconductor}.--- As a next step, we investigate the joint effect of the nonlocal pairing and the proximity induced superconductivity on the spectral properties and the Meissner response of 2DEG placed in contact with a thick SC layer. Our goal here is to demonstrate that one can obtain a nontrivial behavior of the density of states in 2DEG along with the paramagnetic contribution to the Meissner response in a model configuration, which is close to the experiments of Mota and co-workers~\cite{Mota}.
The superconductor is described by the term
\begin{eqnarray}
 H_s = \int d^3\mathbf{R}\biggl[\psi_{s\sigma}^{\dagger}(\mathbf{R})\hat{\xi}_s\psi_{s\sigma}(\mathbf{R}) \\
 \nonumber
 + \Delta_s(\mathbf{R})\psi^{\dagger}_{s\uparrow}(\mathbf{R})\psi_{s\downarrow}^{\dagger}(\mathbf{R}) 
 + \Delta_s^*(\mathbf{R})\psi_{s\downarrow}(\mathbf{R})\psi_{s\uparrow}(\mathbf{R})\biggl] \ ,
\end{eqnarray}
where $\mathbf{R}$ is a three-dimensional vector in the superconducting region, $\hat{\xi}_s = -\nabla_{\mathbf{R}}^2/2m - \mu_s$, and $\Delta_s(\mathbf{R})$ is the superconducting gap function. We choose the creation and annihilation operators in 2DEG $\psi^{\dagger}_{n\sigma}(\mathbf{r})$ and $\psi_{n\sigma}(\mathbf{r})$ to be normalized to the layer volume $\{\psi_{n\sigma}(\mathbf{r}),\psi_{n\sigma'}^{\dagger}(\mathbf{r}')\}=d^{-1}\delta_{\sigma\sigma'}\delta(\mathbf{r}-\mathbf{r}')$, where $\mathbf{r} = (X, Y, Z = 0)$, $\{A,B\} = AB + BA$, and $d$ is the thickness of the 2D layer. Up to the factor of $d$, the Hamiltonian of 2DEG, the tunnel Hamiltonian and the interlayer interaction have the form~(\ref{normal_state_Hamiltonian_two_layers}), (\ref{tunneling_Hamiltonian_two_layers}), and (\ref{interlayer_interaction_Hamiltonian}), respectively, with $\psi_{1\sigma}(\mathbf{r})\to \psi_{s\sigma}(\mathbf{r})$ and $\psi_{2\sigma}(\mathbf{r})\to\psi_{n\sigma}(\mathbf{r})$.

Neglecting the effects of the interlayer interaction in the SC layer, we derive the Gor'kov equations for the Matsubara Green's functions in 2DEG $\check{G}_n$. The resulting equations can be significantly simplified when the characteristic interatomic distance in the SC layer $a_0$ is much less than the one in 2DEG~\cite{KopninPRB2011}. Under this model assumption we get a set of local equations~\cite{summplemental}
\begin{subequations}\label{Gorkov_equations_2D_layer_local_main_text}
 \begin{align}
 \left[-i\omega_n + \check{\tau}_z\xi_n(\mathbf{r}_1)-\check{\Sigma}(\omega_n)\right]\check{G}_n(\mathbf{r}_1,\mathbf{r}_2) = \\
 \nonumber
 d^{-1}\delta(\mathbf{r}_1 - \mathbf{r}_2) \ ,\\
 \label{self_energy_local_main_text}
  \check{\Sigma}(\omega_n) = \pi d \nu_s a_0^2\check{t}^{\dagger}\check{g}_s(\omega_n)\check{t} \ ,
 \end{align}
\end{subequations}
where $\nu_s$ is the density of states per spin projection in the normal-metal state of the superconductor, and
\begin{equation}\label{quasiclassical_GF_SC_layer_main_text}
 \check{g}_s(\omega_n)= \frac{i\omega_n -|\Delta_s|\hat{\sigma}_y\check{\tau_y}}{\sqrt{\omega_n^2 + |\Delta_s|^2}} \ ,
\end{equation}
is the quasiclassical Green's function in the SC layer.

Contrary to the previous setup, the model~(\ref{Gorkov_equations_2D_layer_local_main_text}) allows the odd-frequency superconducting correlations in proximitized 2DEG only for the spin-triplet interlayer gap function. We take as a model example
 $\hat{\Delta}_{\rm int}  = d_t\hat{\sigma}_x$.
For simplicity we choose $t$ and $d_t$ to be real numbers. It is convenient to absorb the dimensional prefactors in the self-energy~(\ref{self_energy_local_main_text}) into the definitions of $t$ and $d_t$: $\pi d \nu_st^2a_0^2 \to t^2$, $\pi d \nu_sd_t^2a_0^2  \to d_t^2$, $\pi d \nu_std_ta_0^2 \to td_t$, so that $t^2$, $d_t^2$, and $td_t$ in further consideration are given in the energy units.

Let us, first, discuss the structure of the resulting self-energy in the particle-hole space $\hat{\Sigma}_{ij}$ ($i,j = 1,2$) 
\begin{subequations}\label{self_energy_triplet_explicit_main_text}
 \begin{align}
  \label{self_energy_triplet_diagonal_main_text}
  \hat{\Sigma}_{11} = \frac{i\omega_n(t^2 + d_t^2)}{\sqrt{\omega_n^2 + |\Delta_s|^2}} + \frac{2td_t\Delta_s\hat{\sigma}_z}{\sqrt{\omega_n^2 + |\Delta_s|^2}} \ ,\\
  \label{self_energy_offdiagonal_main_text}  
  \hat{\Sigma}_{12} = \frac{-\Delta_s\left(t^{2} + d_t^2\right)(i\hat{\sigma}_y)}{\sqrt{\omega_n^2 + |\Delta_s|^2}} - \frac{2i\omega_ntd_t\hat{\sigma}_x}{\sqrt{\omega_n^2 + |\Delta_s|^2}} \ . 
 \end{align}
\end{subequations}
The remaining components $\hat{\Sigma}_{22}$ and $\hat{\Sigma}_{21}$ can be obtained from Eqs.~(\ref{self_energy_triplet_explicit_main_text}) via the relations $\hat{\Sigma}_{22}(\omega_n) = -\hat{\Sigma}_{11}^{\rm T}(-\omega_n)$, $\hat{\Sigma}_{21}(\omega_n) = \hat{\Sigma}_{12}^{\dagger}(-\omega_n)$.
The first term in the right-hand side of Eq.~(\ref{self_energy_offdiagonal_main_text}) indicates that the spin-triplet interlayer pairing leads to the enhancement of the spin-singlet superconducting correlations in 2DEG, which survive in the limit $t\to 0$. The second term in Eq.~(\ref{self_energy_offdiagonal_main_text}) shows that in the presence of both tunneling and the interlayer pairing 2DEG features spin-triplet odd-frequency superconducting correlations. The diagonal elements $\hat{\Sigma}_{11}$ and $\hat{\Sigma}_{22}$ 
contain the Zeeman-type terms $\propto td_t\Delta_s\hat{\sigma}_z$, so that the spin-triplet interlayer pairing can also result in an additional spin splitting for quasiparticle states in 2DEG.  

\begin{figure}[htpb]
\centering
\includegraphics[scale = 0.38]{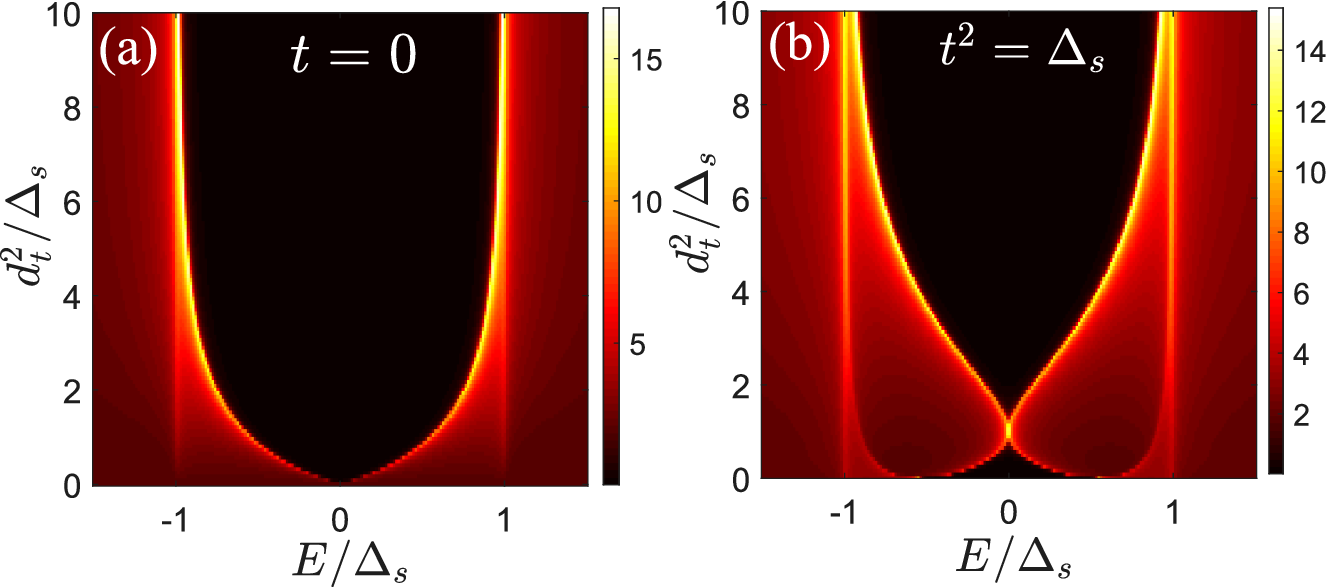}
\caption{Colorplots of the density of states for the model~(\ref{Gorkov_equations_2D_layer_local_main_text}) as a function of energy $E$ and the spin-triplet interlayer gap function $d_t$ for several tunneling rates $t^2/\Delta_s = 0$ (a) and 1 (b). Here $\Delta_s$ is the gap function in the superconducting layer, the model parameters $t^2$ and $d_t^2$ are given in energy units (see Eq.~(\ref{self_energy_local_main_text}) and the corresponding discussion), and we choose the energy level broadening parameter $\Gamma = 0.01\Delta_s$. }
\label{Fig:dos_figure_modified}
\end{figure}

\begin{figure*}[htpb]
\centering
\includegraphics[scale = 0.6]{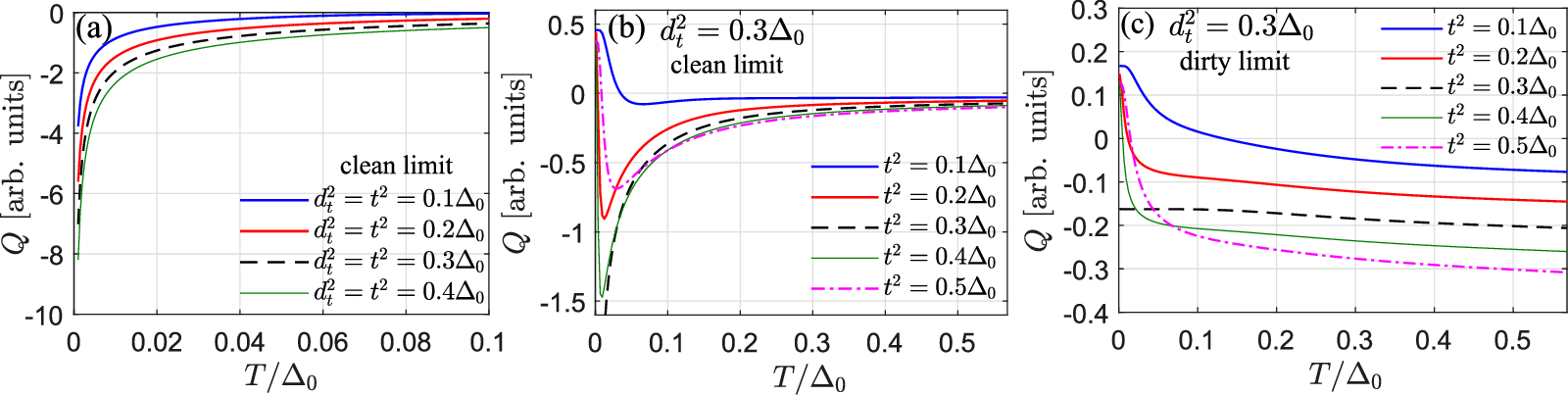}
\caption{Typical $Q(T)$ dependencies. Panels (a) and (b) show the contribution of clean proximitized 2DEG whereas the plots in (c) refer to the dirty limit. We take $t^2 = d_t^2 = 0.1\Delta_0$, $0.2\Delta_0$, $0.3\Delta_0$, and $0.4\Delta_0$ for (a). We choose $d_t^2 = 0.3\Delta_0$ and $t^2/\Delta_0 = 0.1$, 0.2, 0.3, 0.4, and 0.5 for (b) and (c). 
$d_t$ is chosen to be temperature independent and we use the standard interpolation formula for $\Delta_s(T)$ dependence~\cite{GrossZFP1986}. 
Here $\Delta_0 = \Delta_s(T = 0)$.}
\label{Fig:ns_illustration_clean_limit_main_text}
\end{figure*}

We reveal the effects of the spin-triplet interlayer pairing on the spectral properties of proximitized 2DEG by calculating the energy-resolved density of states in the two-dimensional layer~\cite{summplemental}
\begin{equation}\label{2DEG_DOS_definition}
 \nu_{\rm 2D}(E) = \frac{1}{\pi}{\rm Im}{\rm Tr}\left[\hat{G}_n(\mathbf{r},\mathbf{r}; \omega_n \to -iE + \Gamma )\right] \ ,
\end{equation}
where $\Gamma$ is the broadening parameter. Typical colorplots of the density of states are shown in Fig.~\ref{Fig:dos_figure_modified}. Fig.~\ref{Fig:dos_figure_modified}(a) refers to the case $t = 0$ and illustrates the appearance of the induced gap in the spectrum of 2DEG in the absence of tunneling. The density of states for a finite tunneling rate is shown in Fig.~\ref{Fig:dos_figure_modified}(b). The resulting $\nu_{2D}(E)$ dependencies for rather small and fixed $d_t$ are typical for proximitized 2DEG and possess two pair of peaks, one of which marks the induced hard gap in the quasiparticle energy spectrum and another one is located near the gap of the parent superconductor $E\approx \pm \Delta_s$. The increase in the interlayer gap function $d_t$ leads to the splitting of the peaks at the induced gap, which merge into a pronounced zero-bias peak at $d_t = t$. The spectral gap reopens upon further increase in $d_t$ and tends to $2\Delta_s$.

We demonstrate that the induced odd-frequency correlations can provide a paramagnetic contribution to the Meissner response.
For this purpose, we derive linear relations $\mathbf{j} = -Q\mathbf{A}$ between the supercurrent $\mathbf{j}$ and the vector potential $\mathbf{A}$ for the model~(\ref{Gorkov_equations_2D_layer_local_main_text}) within both the clean and dirty limit~\cite{summplemental}. Note that we analyze here only the contribution of the two-dimensional subsystem to the total response while the full response of the structure should be, of course, diamagnetic. For the derivation we follow the standard approach described in Ref.~\cite{Svidzinski}. Typical $Q(T)$ dependencies for the model~(\ref{Gorkov_equations_2D_layer_local_main_text}) with the spin-triplet interlayer pairing are shown in Fig.~\ref{Fig:ns_illustration_clean_limit_main_text}. For simplicity $d_t$ is chosen to be temperature independent and we use the standard interpolation formula
 $\Delta_s(T) = \Delta_0\tanh\left(1.74\sqrt{T_c/T - 1}\right)$ 
for the gap function in the SC layer~\cite{GrossZFP1986} where $T_c$ is the critical temperature of the parent superconductor. For the clean limit and $d_t = t$ (see Fig.~\ref{Fig:ns_illustration_clean_limit_main_text}(a)) the superconducting correlations in 2DEG exhibit the paramagnetic response ($Q<0$) 
which grows with decreasing temperature. Such behavior is similar to the one observed experimentally in~\cite{Mota} which can indicate the relevance of the considered effects for the analysis of puzzling experimental data. The appearance of the paramagnetic response is consistent with the behavior of the density of states, which yield a zero-bias peak at $t = d_t$. Fig.~\ref{Fig:ns_illustration_clean_limit_main_text}(b) illustrates that changing the model parameters away from the zero-bias anomaly ($d_t\neq t$) restores the diamagnetic response at ultra-low temperatures. The results in Fig.~\ref{Fig:ns_illustration_clean_limit_main_text}(c) refer to the dirty limit and also indicate the paramagnetic contribution to the Meissner response at $d_t = t$ within the considered temperature range. In contrast to the clean limit regime, the paramagnetic response at ultra-low temperatures in the dirty limit is possible for not too small interlayer gap function~\cite{summplemental}.

Finally, let us comment on the relation between the direct and inverse proximity effect in superconductor - normal metal structures. In a standard situation rather high transparency of the barrier between the subsystems implies a strong inverse proximity effect. Our results point out that in the presence of the interlayer pairing this relation can break down, namely the inverse proximity effect can be small whereas experimentally measurable effects of the induced superconducting correlations can be noticeable. Note that some indications of such phenomena have been recently observed in~\cite{PestovFTT2019}.

To sum up, we have studied the manifestations of the interlayer pairing in proximitized heterostructures. Depending on the geometry and dimensionality of the system, we have shown that the interlayer pairing can lead to the appearance of the odd-frequency superconducting correlations, FFLO instability, the paramagnetic contribution to the Meissner response, and the multi-peak structure of the density of states. We believe that the obtained results can be useful both for the analysis of experimental data on proximitized heterostructures and for engineering new types of superconducting states in systems with induced superconductivity. Since the considered mechanism can play a role of the Zeeman field, it can be possible that the related effects can be useful for development of new platforms for topologically protected qubits based on Majorana modes~\cite{NayakRMP2008,DasSarmaNPJQI2015}.

This work was supported by the Russian Science Foundation (Grant No. 20-12-00053).

\begin{widetext}

\subsection*{Supplemental Material: Unconventional superconductivity and paramagnetic Meissner response triggered by nonlocal pairing interaction in proximitized heterostructures}

Here we provide the detailed derivation of the results in the main text and additional numerical data. The Material is organized as follows. In Sec.~\ref{2DEG_SC_model} we present the description of the model for the two-dimensional electron gas (2DEG) coupled to a thick conventional superconductor (SC) in the presence of the interlayer pairing. In Sec.~\ref{Gorkov_equations_derivation} we provide the definition of the Matsubara Green's functions and the derivation of the Gor'kov equations for 2DEG/SC system~(11). In Sec.~\ref{spectral_screening_properties} we provide analytical expressions and additional numerical results for the density of states in proximitized 2DEG and the linear response of the induced superconducting correlations in 2DEG to the external magnetic field both in the clean and dirty limits. In Sec.~\ref{gap_equation_two_layer_model} we present the derivation of the self-consistency equation~(7) for the two-layer model.   

\appendix

\section{Model of 2DEG/SC structure}\label{2DEG_SC_model}

Consider a two-dimensional electron gas ($Z = 0$) proximity coupled to a conventional superconductor ($Z > 0$). Hereafter we use the units $k_B = \hbar = 1$, where $k_B$ is the Boltzmann constant and $\hbar$ is the Planck constant. The Hamiltonian of the system reads
\begin{equation}\label{model_Hamiltonian}
 H = H_s + H_n + H_t + H_{\rm int} \ .
\end{equation}
The first term
\begin{eqnarray}
 H_s = \int d^3\mathbf{R}\biggl[\psi_{s\sigma}^{\dagger}(\mathbf{R})\hat{\xi}_s\Psi_{s\sigma}(\mathbf{R}) + \Delta_s(\mathbf{R})\psi^{\dagger}_{s\uparrow}(\mathbf{R})\psi_{s\downarrow}^{\dagger}(\mathbf{R}) + \Delta_s^*(\mathbf{R})\psi_{s\downarrow}(\mathbf{R})\psi_{s\uparrow}(\mathbf{R})\biggl] \ ,
\end{eqnarray}
describes the s-wave superconductor (SC) and the second term
\begin{equation}
 H_n = d\int d^2\mathbf{r} \ \psi_{n\sigma}^{\dagger}(\mathbf{r})\hat{\xi}_n\psi_{n\sigma}(\mathbf{r}) \ ,
\end{equation}
is the Hamiltonian of 2DEG. Here $\sigma = \uparrow$, $\downarrow$ denotes spin degrees of freedom (summation over repeated spin indices is implied), $d$ is the thickness of the normal-metallic layer, $\hat{\xi}_s = (-i\nabla_{\mathbf{R}}-\frac{e}{c}\mathbf{A}(\mathbf{R}))^2/2m_s - \mu_s$ and $\hat{\xi}_n = (-i\nabla_{\mathbf{r}}-\frac{e}{c}\mathbf{A}(\mathbf{r}))^2/2m_n - \mu_n$ stand for the quasiparticle kinetic energy operators in the SC and 2DEG with respect to the corresponding chemical potentials $\mu_s$ and $\mu_n$, $\mathbf{A}$ is the vector potential, $m_s$ and $m_n$ are the effective masses of the electrons in the subsystems, and $\Delta_s(\mathbf{R})$ is the superconducting gap function in the SC layer. The creation and annihilation operators in 2DEG $\psi_{n\sigma}^{\dagger}(\mathbf{r})$ and $\psi_{n\sigma}(\mathbf{r})$ are normalized to the layer volume $\{\psi_{n\sigma}(\mathbf{r}),\psi_{n\sigma'}^{\dagger}(\mathbf{r}')\} = d^{-1}\delta_{\sigma\sigma'}\delta(\mathbf{r}-\mathbf{r}')$. The tunnel Hamiltonian has the form:
\begin{eqnarray}
 H_t = d\int d^2\mathbf{r} \biggl[\psi_{s\sigma}^{\dagger}(\mathbf{r})t(\mathbf{r})\psi_{n\sigma}(\mathbf{r}) + \psi_{n\sigma}^{\dagger}(\mathbf{r})t^*(\mathbf{r})\psi_{s\sigma}(\mathbf{r})\biggl] \ ,
\end{eqnarray}
where $t(\mathbf{r})$ is the tinneling matrix element and we denote $\psi_{s\sigma}(\mathbf{r}) = \psi_{s\sigma}(X,Y,Z = 0)$ for brevity. We consider the effects of the interlayer electron pairing on the induced superconductivity and the linear response of proximitized 2DEG to the applied magnetic field. Assuming that the interaction is relevant in the vicinity of the SC/2DEG interface, we choose the following form of the interaction 
\begin{equation}\label{interaction_Hamiltonian}
 H_{\rm int} = \frac{U_0}{2} d\int d^2\mathbf{r} \ \psi^{\dagger}_{s\sigma}(\mathbf{r})\psi_{n\sigma'}^{\dagger}(\mathbf{r})\psi_{n\sigma'}(\mathbf{r})\psi_{s\sigma}(\mathbf{r}) \ .
\end{equation}

\section{Definition of the Matsubara Green's functions and derivation of Eqs.~(11) in the main text}\label{Gorkov_equations_derivation}

Throughout the second part of our work we use the following Green's functions:
\begin{subequations}\label{Green_functions_definition}
 \begin{align}
  \check{G}_s(\mathbf{X}_1,\mathbf{X}_2) = \left\langle T_{\tau}\check{\psi}_s(\mathbf{X}_1)\check{\psi}_s^{\dagger}(\mathbf{X}_2)\right\rangle \ ,\\
  \check{G}_n(\mathbf{x}_1,\mathbf{x}_2) = \left\langle T_{\tau}\check{\psi}_n(\mathbf{x}_1)\check{\psi}_n^{\dagger}(\mathbf{x}_2)\right\rangle \ ,\\
  \check{G}_t(\mathbf{X}_1,\mathbf{x}_2) = \left\langle T_{\tau}\check{\psi}_s(\mathbf{X}_1)\check{\psi}_n^{\dagger}(\mathbf{x}_2)\right\rangle \ ,\\
  \check{\mathcal{G}}_t(\mathbf{x}_1,\mathbf{X}_2) = \left\langle T_{\tau}\check{\psi}_n(\mathbf{x}_1)\check{\psi}_s^{\dagger}(\mathbf{X}_2)\right\rangle \ .
 \end{align}
\end{subequations}
Here $\mathbf{x} = (\mathbf{r},\tau)$ and $\mathbf{X}= (\mathbf{R},\tau)$, $\tau$ is the imaginary time variable in the Matsubara technique, $T_{\tau}$ is the time-ordering operator. We define the Nambu spinors $\check{\psi}_s(\mathbf{X})$ and $\check{\psi}_n(\mathbf{x})$ as 
\begin{subequations}
 \begin{align}
  \check{\psi}_s(\mathbf{X}) = [\psi_{s\uparrow}(\mathbf{X}),\psi_{s\downarrow}(\mathbf{X}),\psi^{\dagger}_{s\uparrow}(\mathbf{X}),\psi^{\dagger}_{s\downarrow}(\mathbf{X})]^{\rm T} \ ,\\ \check{\psi}_n(\mathbf{x}) = [\psi_{n\uparrow}(\mathbf{x}), \psi_{n\downarrow}(\mathbf{x}), \psi^{\dagger}_{n\uparrow}(\mathbf{x}), \psi^{\dagger}_{n\downarrow}(\mathbf{x})]^{\rm T} \ .
 \end{align}
\end{subequations}

For brevity below we present the equations of motion for the field operators and the derivation of the Gor'kov equations in the absence of the external magnetic field. We reveal the full equations with the vector potential in the end of this section. For the considered model~(\ref{model_Hamiltonian}), fermionic operators in the SC layer satisfy the equations of motion
\begin{eqnarray}
 \frac{\partial}{\partial \tau}\check{\psi}_s(\mathbf{X}) = -\left[\check{\tau}_z\xi_s(\mathbf{R}) + \check{\Delta}_s(\mathbf{R})\right]\check{\psi}_s(\mathbf{X}) - d\delta(Z)\check{t}_{\rm tun}(\mathbf{r})\check{\psi}_n(\mathbf{x})  - \frac{U_0}{2}d\delta(Z)\check{\tau}_z\left[\psi_{n\sigma}^{\dagger}(\mathbf{x})\psi_{n\sigma}(\mathbf{x})\right]\check{\psi}_s(\mathbf{X}) \ ,
\end{eqnarray}
where $\check{\tau}_i$ ($i = x,y,z$) are the Pauli matrices acting in the particle-hole space. The tunneling matrix in the Nambu space is defined as $\check{t}_{\rm tun}(\mathbf{r}) = {\rm diag}[t(\mathbf{r}),-t^*(\mathbf{r})]$, and the superconducting gap matrix has the form
\begin{equation}
 \check{\Delta}_s(\mathbf{R}) = \begin{bmatrix}0&\hat{\Delta}_s(\mathbf{R})\\ \hat{\Delta}_s^{\dagger}(\mathbf{R})&0\end{bmatrix} \ ,
\end{equation}
where $\hat{\Delta}_s(\mathbf{R}) = (i\hat{\sigma}_y)\Delta_s(\mathbf{R})$, $\hat{\sigma}_i$ ($i = x,y,z,$) are the Pauli matrices acting in the spin space.
Equations of motion for the field operators in 2DEG are as follows:
\begin{eqnarray}\label{EOM_2DEG}
 \frac{\partial}{\partial \tau}\check{\psi}_n(\mathbf{x}) = -\check{\tau}_z\xi_n(\mathbf{r})\check{\psi}_n(\mathbf{x}) - \check{t}^*_{\rm tun}(\mathbf{r})\check{\psi}_s(\mathbf{x}) -\frac{U_0}{2}\check{\tau}_z\left[\psi_{s\sigma}^{\dagger}(\mathbf{x})\psi_{s\sigma}(\mathbf{x})\right]\check{\psi}_n(\mathbf{x}) \ .
\end{eqnarray}
For the derivation of the Gor'kov equations for the Green's functions~(\ref{Green_functions_definition}) one should decouple thermodynamic averages of four fermionic operators~\cite{AGDbook}. For instance, considering the equation for the normal correlation function in 2DEG $[\hat{G}_n]_{\alpha\beta}$, we have the combinations
\begin{eqnarray}\label{decoupling_scheme}
 \langle T_{\tau}\psi_{s\sigma}^{\dagger}(\mathbf{x}_1)\psi_{s\sigma}(\mathbf{x}_1)\psi_{n\alpha}(\mathbf{x}_1)\psi_{n\beta}^{\dagger}(\mathbf{x}_2) \rangle = \langle T_{\tau} \psi_{s\sigma}^{\dagger}(\mathbf{x}_1)\psi_{s\sigma}(\mathbf{x}_1)\rangle\langle T_{\tau}\psi_{n\alpha}(\mathbf{x}_1)\psi_{n\beta}^{\dagger}(\mathbf{x}_2)\rangle \\
 \nonumber
 -\langle T_{\tau}\psi_{s\sigma}^{\dagger}(\mathbf{x}_1)\psi_{n\alpha}(\mathbf{x}_1)\rangle\langle T_{\tau}\psi_{s\sigma}(\mathbf{x}_1)\psi_{n\beta}^{\dagger}(\mathbf{x}_2)\rangle  +\langle T_{\tau}\psi_{s\sigma}^{\dagger}(\mathbf{x}_1)\psi_{n\beta}^{\dagger}(\mathbf{x}_2)\rangle\langle T_{\tau}\psi_{s\sigma}(\mathbf{x}_1)\psi_{n\alpha}(\mathbf{x}_1)\rangle \ .
\end{eqnarray}
In the present work we focus on the effects of the interlayer anomalous averages represented, for instance, by the last term in the right-hand side of Eq.~(\ref{decoupling_scheme}) and neglect the other contributions. Using Eqs.~(\ref{Green_functions_definition}) and (\ref{EOM_2DEG}), we derive the equations for the Matsubara Green's functions in 2DEG
\begin{eqnarray}\label{Gorkov_equation_2D_imaginary_time}
 \left[\frac{\partial}{\partial \tau_1} + \check{\tau}_z\xi_n(\mathbf{r}_1)\right]\check{G}_n(\mathbf{x}_1,\mathbf{x}_2) + \begin{bmatrix}t^*(\mathbf{r}_1)&\hat{\Delta}_{\rm int}(\mathbf{r}_1)\\ -\hat{\Delta}_{\rm int}^*(\mathbf{r}_1)&-t(\mathbf{r}_1)\end{bmatrix}\check{G}_t(\mathbf{x}_1,\mathbf{x}_2) =  d^{-1}\delta(\mathbf{x}_1 - \mathbf{x}_2) \ ,
\end{eqnarray}
where we introduced the interlayer gap function
\begin{equation}
 \left[\hat{\Delta}_{\rm int}(\mathbf{r})\right]_{\alpha\beta} = -\frac{U_0}{2}\langle \psi_{n\alpha}(\mathbf{x})\psi_{s\beta}(\mathbf{x})\rangle \ .
\end{equation}
Equations for the Green's functions can be more conveniently in the Matsubara frequency representation $\omega_n = 2\pi T(n+1/2)$. We set $\tau = \tau_1 - \tau_2$ and write
\begin{equation}
 \check{G}(\mathbf{r}_1, \mathbf{r}_2) = \int_0^{1/T}d\tau \ \check{G}(\mathbf{x}_1,\mathbf{x}_2)e^{i\omega_n\tau} \ ,
\end{equation}
omitting the frequency argument for brevity. Equations for the Green's functions in 2DEG written in the Matsubara frequency-coordinate representation have the form
\begin{eqnarray}\label{Gorkov_equations_2D_appendix}
 \left[-i\omega_n + \check{\tau}_z\xi_n(\mathbf{r}_1)\right]\check{G}_n(\mathbf{r}_1,\mathbf{r}_2) +  \begin{bmatrix}t^*(\mathbf{r}_1)&\hat{\Delta}_{\rm int}(\mathbf{r}_1)\\ -\hat{\Delta}_{\rm int}^*(\mathbf{r}_1)&-t(\mathbf{r}_1)\end{bmatrix}\check{G}_t(\mathbf{r}_1,\mathbf{r}_2) =  d^{-1}\delta(\mathbf{r}_1 - \mathbf{r}_2) \ .
\end{eqnarray}
We derive the equation for the tunneling Green function in a similar fashion
\begin{eqnarray}\label{Gorkov_equations_mixed_function}
 \left[-i\omega_n + \check{\tau}_z\xi_s(\mathbf{R}_1) + \check{\Delta}_s(\mathbf{R}_1)\right]\check{G}_t(\mathbf{R}_1,\mathbf{r}_2) + d\delta(Z_1)\begin{bmatrix}t(\mathbf{r}_1)&-\hat{\Delta}_{\rm int}^{\rm T}(\mathbf{r}_1)\\ \hat{\Delta}_{\rm int}^{\dagger}(\mathbf{r}_1)&-t^*(\mathbf{r}_1)\end{bmatrix}\check{G}_n(\mathbf{r}_1,\mathbf{r}_2) = 0 \ .
\end{eqnarray}
Neglecting the back action of 2DEG on the superconductor and the effects of the interlayer interaction in the SC layer, the Gor'kov equations in the SC layer read 
\begin{equation}
 \left[-i\omega_n + \check{\tau}_z\xi_s(\mathbf{R}_1) + \check{\Delta}_s(\mathbf{R}_1)\right]\check{G}_s(\mathbf{R}_1,\mathbf{R}_2) = \delta(\mathbf{R}_1 - \mathbf{R}_2) \ .
\end{equation}
To obtain a closed system of equations for the Green's functions in 2DEG we follow Ref.~\cite{KopninPRB2011_appendix} and write the solution of Eq.~(\ref{Gorkov_equations_mixed_function})
\begin{eqnarray}\label{mixed_Green_function_solution}
 \check{G}_t(\mathbf{R}_1,\mathbf{r}_2) = -d\int d^2\mathbf{r}'\check{G}_s(\mathbf{R}_1,\mathbf{r}')\check{t}(\mathbf{r}')\check{G}_n(\mathbf{r}',\mathbf{r}_2) \ .
\end{eqnarray}
Substituting Eq.~(\ref{mixed_Green_function_solution}) into Eq.~(\ref{Gorkov_equations_2D_appendix}) and restoring the vector potential, we get
\begin{subequations}\label{Gorkov_equations_2D_layer_main_text}
 \begin{align}\label{Gorkov_equation_2D_equation}
 \left\{-i\omega_n + \check{\tau}_z\left[\frac{1}{2m_n}\left(-i\nabla_{\mathbf{r}_1}-\check{\tau}_z\frac{e}{c}\mathbf{A}(\mathbf{r}_1)\right)^2-\mu_n\right]\right\}\check{G}_n(\mathbf{r}_1,\mathbf{r}_2)  - \int d^2\mathbf{r} \  \check{\Sigma}(\mathbf{r}_1,\mathbf{r})\check{G}_n(\mathbf{r},\mathbf{r}_2) = d^{-1}\delta(\mathbf{r}_1 - \mathbf{r}_2) \ ,\\
  \check{\Sigma}(\mathbf{r}_1,\mathbf{r}) = d\check{t}^{\dagger}(\mathbf{r}_1)\check{G}_s(\mathbf{r}_1,\mathbf{r})\check{t}(\mathbf{r}) \ .
 \end{align}
\end{subequations}
Here 
$\check{G}_s(\mathbf{r}_1,\mathbf{r})$ stands for the Green's function of an isolated SC layer taken at the SC/2DEG interface $Z_1 = Z = 0$. The 4$\times$4 matrix Green's function in Eqs.~(\ref{Gorkov_equations_2D_layer_main_text}) has the following structure in the particle-hole space
\begin{equation}
 \check{G} = \begin{pmatrix}\hat{G}&\hat{F}\\ \hat{F}^{\dagger}&\hat{\bar{G}}\end{pmatrix} \ .
\end{equation}
In the present work we assume that the in-plane momentum projection is conserved during the tunneling process. In this case the tunneling amplitude $t$ is independent of the coordinate along 2DEG/SC interface~\cite{KopninJETP2013_appendix}. For simplicity, we also assume that the interlayer gap function $\hat{\Delta}_{\rm int}$ is homogeneous. Note that Eqs.~(\ref{Gorkov_equations_2D_layer_main_text}) can be significantly simplified when the characteristic interatomic distance in the SC layer $a_0$ is much less than the one in 2DEG. Indeed, for rapidly oscillating Green's function in the SC layer
\begin{eqnarray}
 \check{G}_s(\mathbf{r}_1,\mathbf{r}) = \frac{m_s}{2\pi}\biggl\{\check{\tau}_z\frac{\cos\left(k_{Fs}|\mathbf{r}_1-\mathbf{r}|\right)}{|\mathbf{r}_1 - \mathbf{r}|} +\check{g}_s(\omega_n)\frac{\sin\left(k_{Fs}|\mathbf{r}_1 - \mathbf{r}|\right)}{|\mathbf{r}_1 - \mathbf{r}|}
 \biggl\}e^{-\frac{m_s\sqrt{\omega_n^2 + |\Delta_s|^2}}{k_{Fs}}|\mathbf{r}_1-\mathbf{r}|} \ ,
\end{eqnarray}
the integral in Eq.~(\ref{Gorkov_equation_2D_equation}) converges at $|\mathbf{r}_1 - \mathbf{r}|\sim a_0$, and the resulting self-energy is local. Thus, under our model assumptions Eqs.~(\ref{Gorkov_equations_2D_layer_main_text}) acquire the form~(11) in the main text.

\section{Spectral and screening properties of 2DEG}\label{spectral_screening_properties}

For the calculations of the density of states~(14), we solve Eq.~(11a) with a local self-energy given by Eqs.~(13). As a first step, we derive the expression for the normal Matsubara Green's function in 2DEG at coincident spatial arguments
\begin{eqnarray}\label{normal_Matsubara_coincident_arguments}
 \left[\hat{G}_n(\mathbf{r},\mathbf{r})\right]_{\sigma\sigma} = \int\frac{d^2\mathbf{k}}{(2\pi)^2}\frac{[i(\tilde{\omega}_n - i\sigma h) + \xi_n]}{\left[\xi_n^2 + (\tilde{\omega}_n - i\sigma h)^2 + f_{\sigma}^2\right]} \ ,
\end{eqnarray}
where $\sigma = \uparrow,\downarrow$ ($\pm 1$), $\xi_n = \mathbf{k}^2/2m_n - \mu_n$, and
\begin{equation}\label{normal_Matsubara_functions}
  \tilde{\omega}_n = \omega_n\left(1 + \frac{t^2 + d_t^2}{\sqrt{\omega_n^2 + \Delta_s^2}}\right) \ , \ \
  h(\omega_n) = \frac{2td_t\Delta_s}{\sqrt{\omega_n^2 + \Delta_s^2}} \ , \ \ 
  f_{\sigma}(\omega_n) = \frac{\Delta_s(t^2 + d_t^2) + 2i\sigma \omega_n td_t}{\sqrt{\omega_n^2 + \Delta_s^2}} \ .
\end{equation}
Integration over the momentum in Eq.~(\ref{normal_Matsubara_coincident_arguments}) yields
\begin{equation}\label{normal_Matsubara_quasiclassical}
 \left[\hat{G}_n(\mathbf{r},\mathbf{r})\right]_{\sigma\sigma} = \pi\nu_0\frac{i\tilde{\omega}_n + \sigma h}{\sqrt{f_{\sigma}^2 - (i\tilde{\omega}_n + \sigma h)^2}} \ .
\end{equation}
Here $\nu_0 = m_n/2\pi$ is the density of states in an isolated 2DEG per spin projection. Finally, we substitute Eqs.~(\ref{normal_Matsubara_quasiclassical}) into Eq.~(14) and calculate the density of states.

\begin{figure*}[htpb]
\centering
\includegraphics[scale = 0.5]{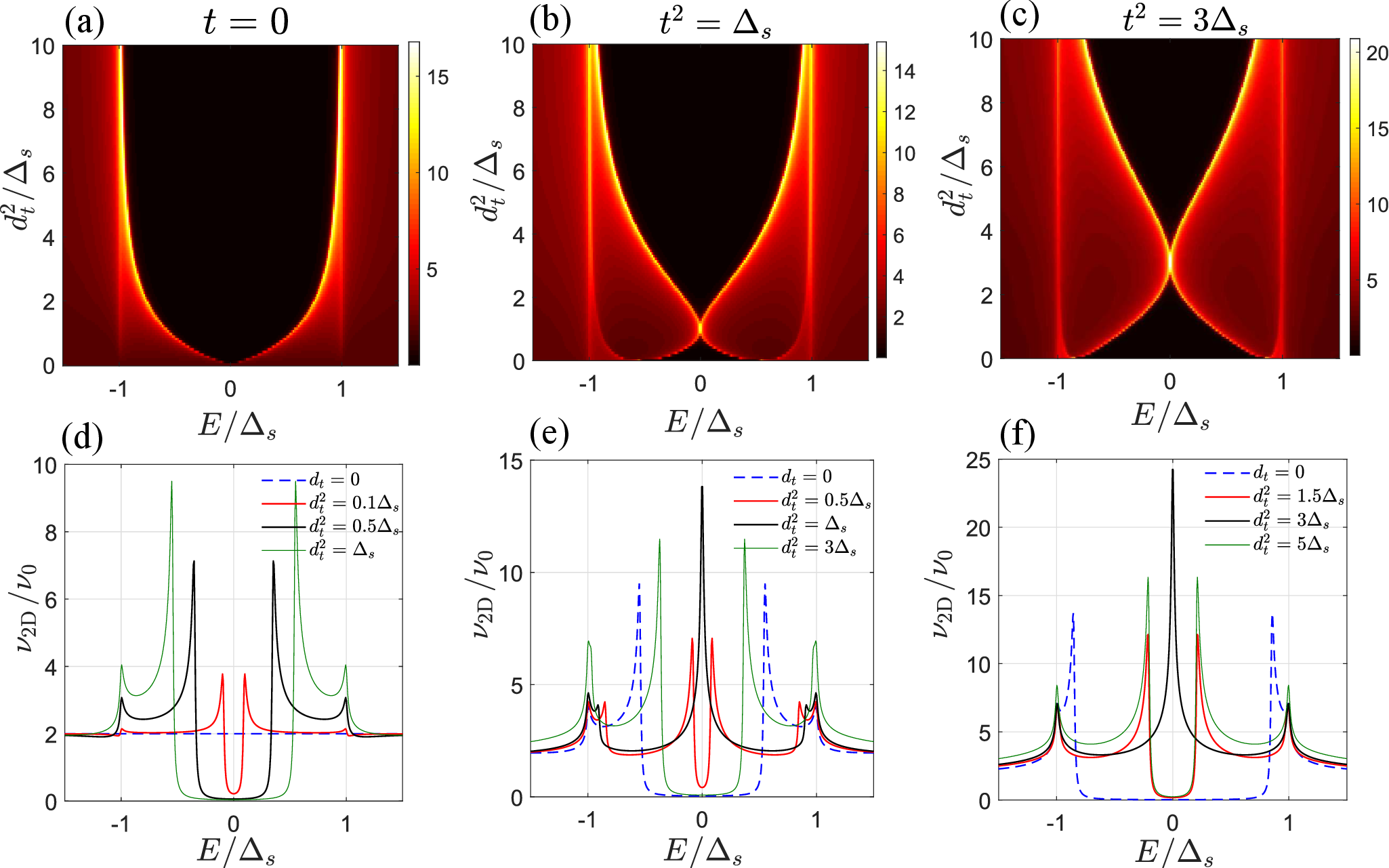}
\caption{Typical energy dependencies of the density of states in the two-dimensional layer $\nu_{\rm 2D}$. (a)-(c) Colorplots of the density of states as a function of energy $E$ and the interlayer gap function $d_t$ for several values of the tunneling rate $t^2$. (d)-(f) $\nu_{\rm 2D}(E)$ dependencies for various values of the interlayer gap function $d_t$. Panels (a) and (d), (b) and (e), and (c) and (f) correspond to $t = 0$, $t^2 = \Delta_s$, and $t^2 = 3\Delta_s$, respectively. We choose the energy level broadening parameter $\Gamma = 0.01\Delta_s$ to produce the plots.}
\label{Fig:dos_figure1}
\end{figure*}

Typical behavior of the density of states in 2DEG as a function of energy and model parameters are presented in Fig.~\ref{Fig:dos_figure1}. Panels (a)-(c) show the colorplots of the density of states as a function of energy $E$ and the interlayer gap function $d_t$ for several tunneling rates $t^2 = 0$, $t^2=\Delta_s$, and $t^2 = 3\Delta_s$, respectively. Panels (d)-(f) reveal $\nu_{\rm 2D}(E)$ plots for several values of the interlayer gap function. We choose the energy level broadening parameter $\Gamma = 0.01\Delta_s$ to produce the plots. Figs.~\ref{Fig:dos_figure1}(a) and~\ref{Fig:dos_figure1}(d) refer to the case $t = 0$, for which the two-dimensional layer only features the spin-singlet superconducting correlations [see Eq.~(13b)]. One can see the emerging minigap in the density of states for rather small $d_t$ values (see the solid red line in Fig.~\ref{Fig:dos_figure1}(d)). The magnitude of the minigap for $d_t^2 = 0.1\Delta_s$ is approximately $0.2\Delta_s$, which is in agreement with the result of Eq.~(13b) in the case $t = 0$ and $d_t^2\ll \Delta_s$. Two additional features in the density of states are located near the energy gap of the parent superconductor $E \approx \pm \Delta_s$. The colorplot in Fig.~\ref{Fig:dos_figure1}(a) shows that the spectral gap tends to $2\Delta_s$ upon the increase in the absolute value of the interlayer gap function. We provide $\nu_{\rm 2D}(E)$ plots for a finite tunneling rate $t^2 = \Delta_s$ and $d_t^2/\Delta_s = 0$, $0.5$, $1$, and $3$ in Fig.~\ref{Fig:dos_figure1}(e). Corresponding $\nu_{\rm 2D}(E)$ curve for $d_t = 0$ (shown by a blue dashed line) represents a typical energy dependence of the density of states of 2DEG with the induced superconductivity and possesses two pair of peaks, one of which (at $E\approx \pm 0.55\Delta_s$) marks the induced hard gap in the energy spectrum and another one is located at $E\approx \pm \Delta_s$. The increase in the interlayer gap function leads to the splitting of the peaks at the induced gap and to the decrease in the induced gap, which eventually disappears at a certain value of interlayer pairing amplitude. The black solid line in Fig.~\ref{Fig:dos_figure1}(e) shows a pronounced zero-bias peak in the density of states at $d_t = t$. The colorplot in Fig.~\ref{Fig:dos_figure1}(b) demonstrates that the spectral gap reopens upon further increase in $d_t$ and tends to $2\Delta_s$ for rather large $d_t$ values. The results in Figs.~\ref{Fig:dos_figure1}(c) and~\ref{Fig:dos_figure1}(f) obtained for larger tunneling rate $t^2 = 3\Delta_s$ also demonstrate the hard gap closing-reopening feature upon the variation of the interlayer pairing amplitude as well as the appearance of a zero-bias peak in the density of states of the two-dimensional system at $d_t = t$. 

We continue with the analysis of a linear response of the induced superconducting correlations in 2DEG to an external magnetic field. Corresponding linear relations between the supercurrent $\mathbf{j}$ and the vector potential $\mathbf{A}$ in 2DEG are derived within both the clean and dirty limit. For the derivation we choose the transverse gauge for the vector potential ${\rm div}\mathbf{A} = 0$ and follow the approach described in Ref.~\cite{Svidzinski_appendix}.

\subsection{Clean limit}

Here we consider the ballistic case. For the derivation of the quasiclassical equations in 2DEG, we introduce the Matsubara Green's functions in the mixed representation
\begin{equation}\label{Wigner_transform}
 \check{G}_n(\mathbf{r},\mathbf{k}) = \int d\delta\mathbf{r} \ e^{-i\mathbf{k}\delta\mathbf{r}}\check{G}_n(\mathbf{r},\delta\mathbf{r}) \ ,
\end{equation}
where $\mathbf{r} = (\mathbf{r}_1 + \mathbf{r}_2)/2$ and $\delta\mathbf{r} = \mathbf{r}_1 - \mathbf{r}_2$. Using Eqs.~(\ref{Gorkov_equation_2D_equation}), (\ref{Wigner_transform}) and considering the quasiparticle states in the vicinity of the Fermi surface
\begin{equation}
 \mathbf{k} = \mathbf{n}(k_{Fn} + \xi_n/v_{Fn}) \ ,
\end{equation}
we derive the quasiclassical equations for the Green's function in the mixed representation.
\begin{eqnarray}\label{Gorkov_equation_mixed_representation}
 \biggl[-i\omega_n + \check{\tau}_z\left(\xi_n - \frac{i}{2}\mathbf{v}_{Fn}\nabla_{\mathbf{r}}\right) -e\mathbf{v}_{Fn}\mathbf{A}\left(\mathbf{r}+\frac{i}{2}\mathbf{v}_{Fn}\frac{d}{d\xi_n}\right)\biggl]\check{G}_n(\mathbf{r},\mathbf{n},\xi_n) -\check{\Sigma}(\mathbf{r})\check{G}_n(\mathbf{r},\mathbf{n},\xi_n) = 1 \ .
\end{eqnarray}
Here $\mathbf{v}_{Fn} = v_{Fn}\mathbf{n}$, $v_{Fn}$ denotes the Fermi velocity in an isolated 2DEG, $\mathbf{n} = [\cos\varphi_{\mathbf{k}},\sin\varphi_{\mathbf{k}},0]$, $k_{Fn} = m_nv_{Fn}$, and $\xi_n$ is the kinetic energy of quasiparticles relative to the chemical potential. Note that in the above equation we used the local approximation for the self-energy.
The supercurrent density is then determined from the solution of Eq.~(\ref{Gorkov_equation_mixed_representation}) 
\begin{equation}\label{supercurrent_definition}
 \mathbf{j}(\mathbf{r}) = -ek_{Fn}T\sum_{\omega_n}\int\frac{d\xi_n}{(2\pi)}\frac{d\mathbf{n}}{(2\pi)}\mathbf{n}{\rm Tr}\left[\hat{G}_n(\mathbf{r},\mathbf{n},\xi_n)\right] \ .
\end{equation}

As a next step, we find the first-order correction for the Green's function with respect to the vector potential. For this purpose, it is convenient to calculate the Fourier transform of the Green's function with respect to $\xi_n$
\begin{equation}\label{q_representation}
 \check{G}_n(q) = \int \check{G}_n(\xi_n)e^{iq\xi_n}\frac{d\xi_n}{2\pi} \ .
\end{equation} 
Using Eq.~(\ref{Gorkov_equation_mixed_representation}), we derive the quasiclassical equation for the Fourier transform~(\ref{q_representation}). Eliminating the spatial derivative via the replacement $\mathbf{r}\to \mathbf{r} - \frac{1}{2}\mathbf{v}_{Fn}q$, we get the equation
\begin{eqnarray}\label{quasiclassical_fourier_space}
 \biggl[-i\omega_n - i\check{\tau}_z\frac{\partial}{\partial q} - e\mathbf{v}_{Fn}\mathbf{A}\left(\mathbf{r} + q\mathbf{v}_{Fn}\right)\biggl]\check{G}_n(q) - \check{\Sigma}\left(\mathbf{r} + \frac{1}{2}q\mathbf{v}_{Fn}\right)\check{G}_n(q) = \delta(q) \ .
\end{eqnarray}
Note that for the derivation of the linear response it is sufficient to expand the Green's function up to the first-order term in the vector potential 
\begin{equation}\label{perturbation_expansion}
  \check{G}_n(q)\approx \check{G}_n^{(0)}(q) + \check{G}_n^{(1)}(q) \ , 
\end{equation}
and take the unperturbed homogeneous self-energy $\check{\Sigma}^{(0)}$. Indeed, within the local approximation the self-energy involves the Green's function in the superconducting layer at coincident spatial arguments, so the first-order correction $\check{\Sigma}^{(1)}$ should vanish upon averaging over the momentum directions. Unperturbed Green's functions have the form
\begin{equation}\label{zero_order_solutions}
  \hat{G}_n^{(0)}(q) = \sum_{\sigma = \uparrow,\downarrow}\hat{\Pi}_{z\sigma}G^{(0)}_{n\sigma}(q) \ , \ \ \
  \hat{F}_n^{\dagger(0)}(q) = -(i\hat{\sigma}_y)\sum_{\sigma = \uparrow,\downarrow}\hat{\Pi}_{z\sigma}F_{n\sigma}^{\dagger(0)}(q) \ ,
\end{equation}
where 
$\hat{\Pi}_{z\uparrow,\downarrow} = (1 \pm \hat{\sigma}_z)/2$ and the expressions for the components read as
\begin{equation}\label{zero_order_components}
  G_{n\sigma}^{(0)}(q) = \frac{\gamma_{\sigma}(q)}{2}\left[\frac{i\tilde{\omega}_n + \sigma h}{\sqrt{f_{\sigma}^2 - (i\tilde{\omega}_n + \sigma h)^2}}+ i{\rm sgn}(q)\right] \ , \ \ 
  F_{n\sigma}^{\dagger(0)}(q) = \frac{\gamma_{\sigma}(q)}{2}\frac{f_{\sigma}}{\sqrt{f_{\sigma}^2-(i\tilde{\omega}_n+\sigma h)^2}} \ ,
\end{equation}
with $\gamma_{\sigma}(q) = \exp[-\sqrt{f_{\sigma}^2 - (i\tilde{\omega}_n + \sigma h)^2}|q|]$. 
The other Green's functions $\hat{\bar{G}}_n^{(0)}$ and $\hat{F}_n^{(0)}$ can be obtained from Eqs.~(\ref{zero_order_solutions}) by using the symmetry relations $\hat{\bar{G}}_n^{(0)}(\omega_n) = -\hat{G}_n^{(0)}(-\omega_n)$ and $\hat{F}_n^{(0)}(\omega_n) = [\hat{F}_n^{\dagger(0)}(\omega_n)]^{\rm T}$. The first-order correction for the Green's function at $q = 0$ is determined from the expression
\begin{equation}\label{first_order_correction}
 \check{G}_n^{(1)}(q = 0) = \int dq' \check{G}_n^{(0)}(-q')e\mathbf{v}_{Fn}\mathbf{A}(\mathbf{r} + q'\mathbf{v}_{Fn})\check{G}_n^{(0)}(q') \ .
\end{equation}
We put
\begin{equation}
  \mathbf{A}(\mathbf{r}) = \int \frac{d^2\mathbf{k}}{(2\pi)^2} \ \mathbf{A}(\mathbf{k})e^{i\mathbf{k}\mathbf{r}} \ , \ \ \
  \mathbf{j}(\mathbf{r}) = \int \frac{d^2\mathbf{k}}{(2\pi)^2} \ \mathbf{j}(\mathbf{k})e^{i\mathbf{k}\mathbf{r}} \ ,
\end{equation}
and then substitute Eqs.~(\ref{zero_order_solutions}) and (\ref{zero_order_components}) into Eq.~(\ref{first_order_correction}). Performing the integration, we derive a linear relation between the supercurrent and vector potential in the clean limit
\begin{eqnarray}
 \mathbf{j}(\mathbf{k}) = -e^2p_{Fn}v_{Fn}T\sum_{\omega_n}\sum_{\sigma = \uparrow,\downarrow}\frac{1}{2}\frac{f_{\sigma}^2}{\sqrt{f_{\sigma}^2 - (i\tilde{\omega}_n + \sigma h)^2}} \int\frac{d\mathbf{n}}{(2\pi)}\frac{\mathbf{n}\left(\mathbf{n}\mathbf{A}(\mathbf{k})\right)}{\left[f_{\sigma}^2 - (i\tilde{\omega}_n + \sigma h)^2 + v_{Fn}^2(\mathbf{n}\mathbf{k})^2/4\right]} \ .
\end{eqnarray}
Under the assumption of a local response, the above equation transforms as follows:
\begin{subequations}\label{jA_relation_clean_limit}
\begin{align}
 \label{local_linear_relation_clean_limit}
 \mathbf{j}(\mathbf{r}) = -Q\mathbf{A}(\mathbf{r}) \ ,\\
 \label{ns_clean_limit}
 Q = e^2\frac{k_{Fn}v_{Fn}}{4}T\sum_{\omega_n,\sigma}\frac{f_{\sigma}^2}{\left[f_{\sigma}^2 - (i\tilde{\omega}_n + \sigma h)^2\right]^{3/2}} \ .
 \end{align}
\end{subequations}

\begin{figure*}[htpb]
\centering
\includegraphics[scale = 0.8]{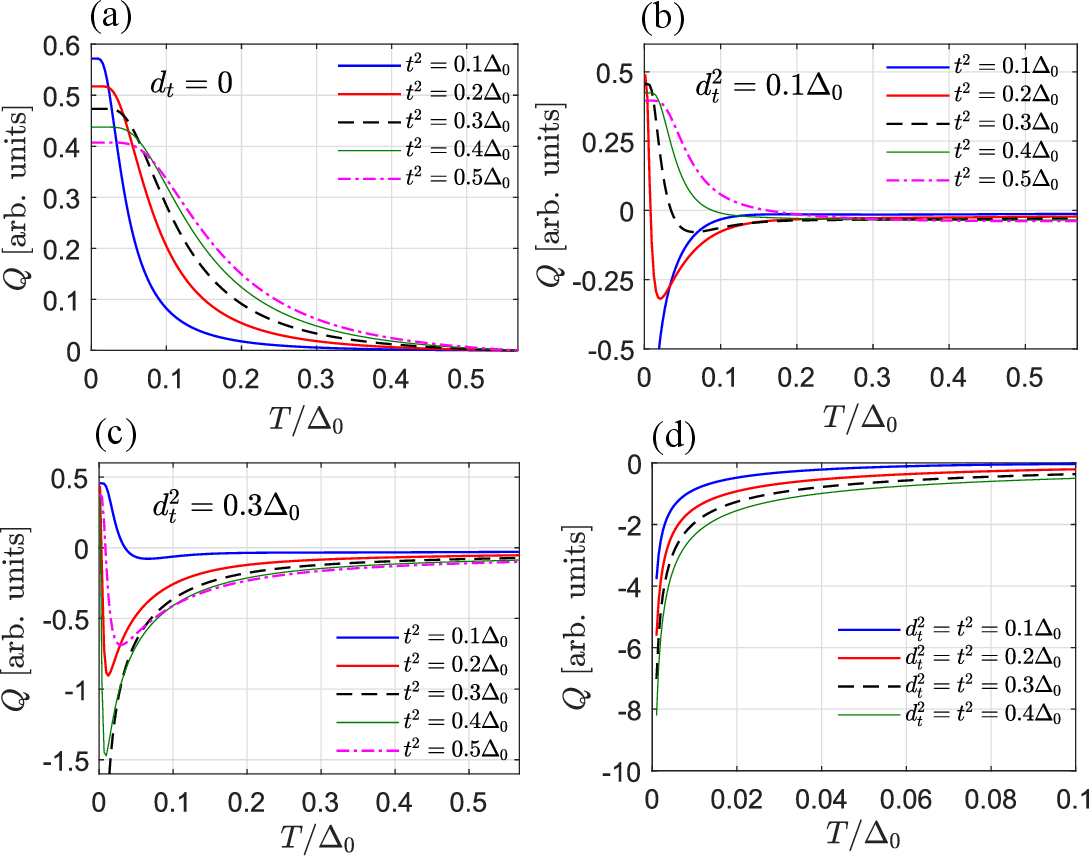}
\caption{Typical temperature dependencies of the coefficient $Q$ in the linear relation between the supercurrent and the vector potential~(\ref{local_linear_relation_clean_limit}). The plots represent the results of Eq.~(\ref{ns_clean_limit}), which is valid in the clean limit and under the assumption of a local response. Panels (a), (b), and (c) correspond to $d_t^2 = 0$, $0.1\Delta_0$, and $0.3\Delta_0$, respectively. (d) Several $Q(T)$ plots for $t^2 = d_t^2 = 0.1\Delta_0$, $0.2\Delta_0$, $0.3\Delta_0$, and $0.4\Delta_0$. The standard interpolation formula was used for $\Delta_s(T)$ dependence whereas we choose the interlayer pairing amplitude $d_t$ to be constant throughout the considered temperature range. Here $\Delta_0 = \Delta_s(T = 0)$.}
\label{Fig:ns_illustration_clean_limit}
\end{figure*} 

Typical temperature dependencies of the coefficient $Q$ in the linear relation~(\ref{jA_relation_clean_limit}) are shown in Fig.~\ref{Fig:ns_illustration_clean_limit}. For simplicity, we choose the interlayer gap function $d_t$ to be constant within the considered temperature range. To take into account the temperature dependence of the gap function in the SC layer, we use the interpolation formula $\Delta_s(T) = \Delta_0\tanh(1.74\sqrt{T_c/T - 1})$, where $\Delta_0 = \Delta_s(T = 0)$ and $T_c$ denotes the critical temperature of the parent superconductor. Figs.~\ref{Fig:ns_illustration_clean_limit}(a), \ref{Fig:ns_illustration_clean_limit}(b), and \ref{Fig:ns_illustration_clean_limit}(c) show several $Q(T)$ plots for a fixed interlayer gap function and several tunneling rates $t^2/\Delta_0 = 0.1$, $0.2$, $0.3$, $0.4$, and $0.5$. The results in Fig.~\ref{Fig:ns_illustration_clean_limit}(a) for $d_t = 0$ indicate that in the absence of the interlayer spin-triplet pairing the induced superconducting correlations in 2DEG only exhibit the Meissner response ($Q>0$). Diamagnetic response of the induced Cooper pairs becomes more pronounced at lower temperatures with decreasing $t^2$. This behavior is consistent with the fact that the induced gap in the quasiparticle energy spectrum of the two-dimensional layer decreases upon the decrease in the tunneling rate. $Q(T)$ plots in Figs.~\ref{Fig:ns_illustration_clean_limit}(b) and~\ref{Fig:ns_illustration_clean_limit}(c) reveal several qualitatively different types of the linear response within different temperature ranges. For $d_t^2 = t^2 = 0.1\Delta_0$ and $0.3\Delta_0$ [see a blue solid line in Fig.~\ref{Fig:ns_illustration_clean_limit}(b) and a black dashed line in Fig.~\ref{Fig:ns_illustration_clean_limit}(c)], the superconducting correlations in 2DEG exhibit the paramagnetic response ($Q<0$) within the considered temperature range, and $|Q|$ grows with decreasing temperature. We note that this behavior is in qualitative agreement with our calculations of the density of states, which yield a zero-bias anomaly at $t = d_t$. The parameter range $t^2 > d_t^2$ ($t^2 < d_t^2$) is characterized by the presence of the minimum on a $Q(T)$ curve and a diamagnetic response at low temperatures. For clarity, we also reveal the low-temperature behavior of $Q$ for $d_t^2 = t^2 = 0.1\Delta_0$, $0.2\Delta_0$, $0.3\Delta_0$, and $0.4\Delta_0$ in Fig.~\ref{Fig:ns_illustration_clean_limit}(d). Note that the presence of the paramagnetic response of 2DEG at high temperatures in Figs.~\ref{Fig:ns_illustration_clean_limit}(b) and~\ref{Fig:ns_illustration_clean_limit}(c) is probably related to the presence of the spin-triplet superconducting correlations in 2DEG, which survive in the limit $\Delta_s \to 0$ .  

\subsection{Dirty limit}

\begin{figure*}[htpb]
\centering
\includegraphics[scale = 0.8]{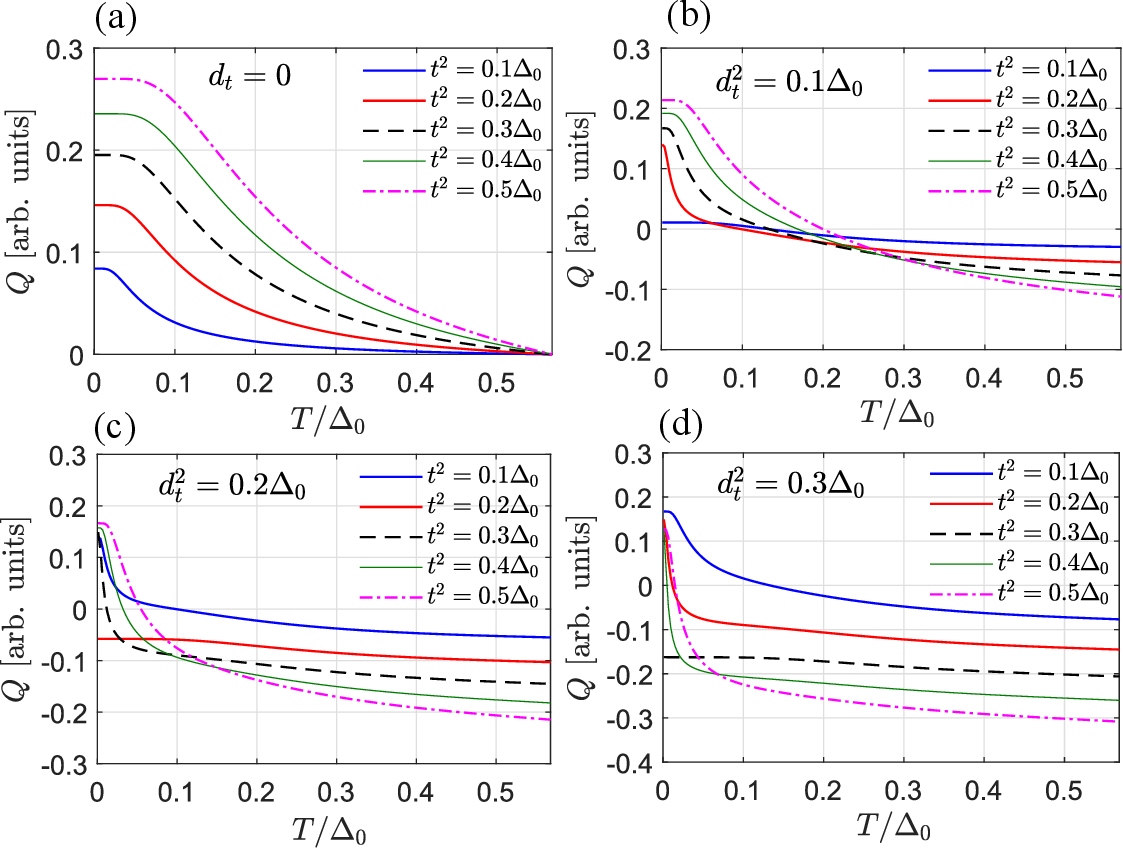}
\caption{Typical temperature dependencies of the coefficient $Q$ in the linear relation~(\ref{linear_relation_dirty_limit}). The plots are the results of Eq.~(\ref{coefficient_dirty_limit}), which is valid in the dirty limit. Panels (a), (b), (c), and (d) correspond to $d_t^2 = 0$, $0.1\Delta_0$, $0.2\Delta_0$ and $0.3\Delta_0$, respectively. The standard interpolation formula was used for $\Delta_s(T)$ dependence whereas we choose the
interlayer pairing amplitude $d_t$ to be constant throughout the considered temperature range.}
\label{Fig:ns_illustration_dirty_limit}
\end{figure*} 

We proceed with the analysis of the linear response of the induced superconducting correlations in 2DEG with randomly distributed nonmagnetic point impurities. The effects of an elastic scattering are described by the impurity self-energy
\begin{equation}
 \check{\Sigma}_{\rm imp}(\mathbf{r}) = \frac{1}{\tau}\int\frac{d\xi_n}{2\pi}\frac{d\mathbf{n}}{2\pi}\check{\tau}_z\check{G}_n(\mathbf{r},\mathbf{n},\xi_n)\check{\tau}_z \ ,
\end{equation}
included into Eq.~(\ref{Gorkov_equation_mixed_representation}). Here $\tau$ is the average time between collisions. As a first step, we derive the Eilenberger equations for $\xi_n$-integrated Green's functions. For this purpose, we subtract Eq.~(\ref{Gorkov_equation_mixed_representation}) and its transpose. As a result, we get 
\begin{equation}\label{Eilenberger_equation}
 -i\mathbf{v}_{Fn}\nabla_{\mathbf{r}}\check{g}_n(\mathbf{r},\mathbf{n}) - \left[\check{w}(\mathbf{r}), \check{g}_n(\mathbf{r},\mathbf{n})\right]  = 0 \ ,
\end{equation}
where 
\begin{equation}
 \check{w} = \check{\tau}_z\left[i\omega_n + \check{\Sigma}(\mathbf{r}) + \check{\Sigma}_{\rm imp}(\mathbf{r}) + e\mathbf{v}_{Fn}\mathbf{A}(\mathbf{r}) \right] \ ,
\end{equation}
and the quasiclassical Green's function is defined as follows:
\begin{equation}\label{quasiclassical_GF_definition}
 \check{g}_n(\mathbf{r},\mathbf{n}) = \int\frac{d\xi_n}{2\pi}\check{G}_n(\mathbf{r},\mathbf{n},\xi_n)\check{\tau}_z \ .
\end{equation}
Using Eqs.~(\ref{q_representation}), (\ref{zero_order_solutions}), and (\ref{zero_order_components}) evaluated at $q = 0$, it is straightforward to show that the introduced quasiclassical Green's function~(\ref{quasiclassical_GF_definition}) obeys the normalization condition $\check{g}_n^2 = -1/4$. In this subsection we consider the case when the mean free path for elastic scattering $\ell$ is much less than the spatial scale of the superconducting correlations in 2DEG. In this case one can seek the solution of Eq.~(\ref{Eilenberger_equation}) in the form
\begin{equation}\label{dirty_limit_GF_expansion}
 \check{g}_n(\mathbf{r},\mathbf{n}) = \check{g}_n^{(0)}(\mathbf{r}) + \mathbf{n}\check{\mathbf{\Gamma}}_n(\mathbf{r}) \ .
\end{equation}
Isotropic part of the Green function $\check{g}_n^{(0)}$ satisfies the Usadel equation
\begin{equation}\label{Usadel_equation}
 D_n\check{\nabla}_{\mathbf{r}}\left[\check{g}_n^{(0)}\check{\nabla}_{\mathbf{r}}\check{g}_n^{(0)}\right] - \frac{1}{2}\left[\check{\tau}_z\left(i\omega_n + \check{\Sigma}\right), \check{g}_n^{(0)}\right] = 0 \ ,
\end{equation}
whereas a small correction $\check{\mathbf{\Gamma}}_n$ is determined from the expression
\begin{equation}\label{dirty_limit_anisotropic_correction}
 \check{\mathbf{\Gamma}}_n(\mathbf{r}) = 2i\ell\check{g}_n^{(0)}(\mathbf{r})\check{\nabla}_{\mathbf{r}}\check{g}^{(0)}_n(\mathbf{r}) \ .
\end{equation}
In the above equations $\check{\nabla}_{\mathbf{r}}\check{a} = \nabla_{\mathbf{r}} \check{a} - ie\mathbf{A}\left[\check{\tau}_z, \check{a}\right]$ and $D_n = v_{Fn}\ell/2$ is the diffusion coefficient in 2DEG. Substituting Eqs.~(\ref{dirty_limit_GF_expansion}) and (\ref{dirty_limit_anisotropic_correction}) into Eq.~(\ref{supercurrent_definition}), we get the expression for the supercurrent
\begin{eqnarray}\label{supercurrent_Usadel}
 \mathbf{j}(\mathbf{r}) = -2ieD_n\nu_nT\sum_{\omega_n}{\rm Tr}\Bigl[\hat{g}_n^{(0)}(\mathbf{r})\nabla_{\mathbf{r}}\hat{g}_n^{(0)}(\mathbf{r}) + \hat{f}_n^{(0)}(\mathbf{r})(\nabla_{\mathbf{r}} + 2ie\mathbf{A}(\mathbf{r}))\hat{f}_n^{\dagger(0)}(\mathbf{r})\Bigl] \ .
\end{eqnarray}
Note that for the chosen gauge of the vector potential ${\rm div}\mathbf{A} = 0$, the Usadel equation~(\ref{Usadel_equation}) doesn't contain linear terms in $\mathbf{A}$. Correspondingly, the linear relation between the supercurrent and the vector potential can be obtained by substituting the zero-order spatially homogeneous Green's functions defined by Eqs.~(\ref{zero_order_solutions}) and (\ref{zero_order_components}) into Eq.~(\ref{supercurrent_Usadel}). As a result, we obtain the local relation
\begin{subequations}
 \begin{align}\label{linear_relation_dirty_limit}
  \mathbf{j}(\mathbf{r}) = -Q\mathbf{A}(\mathbf{r}) \ ,\\
  \label{coefficient_dirty_limit}
  Q = 2\pi e^2D_n\nu_nT\sum_{\omega_n}\sum_{\sigma = \uparrow,\downarrow}\frac{f_{\sigma}^2}{\left[f_{\sigma}^2 - (i\tilde{\omega}_n + \sigma h\right]^2} \ .
 \end{align}
\end{subequations}

Typical temperature dependencies of the coefficient $Q$ in the linear relation~(\ref{linear_relation_dirty_limit}) are shown in Fig.~\ref{Fig:ns_illustration_dirty_limit}. The plots are the results of Eq.~(\ref{coefficient_dirty_limit}). Panels (a), (b), (c), and (d) correspond to $d_t^2/\Delta_0 = 0$, 0.1, 0.2, and 0.3, respectively. Fig.~\ref{Fig:ns_illustration_dirty_limit}(a) shows that in the absence of the spin-triplet interlayer pairing the induced superconducting correlations in 2DEG only exhibit the diamagnetic response. In constrast with the corresponding results for the clean limit [see Fig.~\ref{Fig:ns_illustration_clean_limit}(a)], the plots in Fig.~\ref{Fig:ns_illustration_dirty_limit}(a) demonstrate that the magnitude of the response $|Q|$ at low temperatures grows with increasing tunneling rate. Similarly to the previously considered case, we find that in the case of a finite interlayer gap function the type of the linear response can vary with temperature. In particular, the results for $d_t^2 = t^2 = 0.1\Delta_0$ [shown by a blue solid line in Fig.~\ref{Fig:ns_illustration_dirty_limit}(b)] reveal rather small diamagnetic response at low temperatures, which switches into the paramagnetic one upon the increase in $T$. The increase in the tunneling rate $t^2$ results in the enhancement of both the diamagnetic and paramagnetic response. The temperature range corresponding to the diamagnetic response increases for larger tunneling rates. Panels (c) and (d) show typical $Q(T)$ plots within both parameter regions $t^2<d_t^2$ and $t^2 > d_t^2$. Considering, for instance, Fig.~\ref{Fig:ns_illustration_dirty_limit}(d), we see that for $t^2 = 0.1\Delta_0$ the two-dimensional layer features the diamagnetic (paramagnetic) response at low (high) temperatures. The temperature range, within which the Meissner response is established shrinks upon the increase in $t^2$. At $t^2 = d_t^2 = 0.3\Delta_0$ [see a black dashed line in Fig.~\ref{Fig:ns_illustration_dirty_limit}(d)] 2DEG exhibits the paramagnetic response within the considered temperature range. Further increase in the tunneling rate $t^2>d_t^2$ restores the low-temperature diamagnetic response and also leads to the enhancement of the paramagnetic response at high temperatures.

\section{Derivation of the gap equation~(7) for the two-layer model}\label{gap_equation_two_layer_model}

Here we consider the two-layer model and provide the derivation of the gap equation~(7) in the main text. Our starting point is the Gor'kov equations~(4)  
\begin{subequations}
 \begin{align}
  \left(-i\omega_n + \check{\tau}_z\xi_{2\mathbf{k}}\right)\check{G}_{22} + \check{t}^{\dagger}\check{G}_{12} = 1 \ ,\\
  \left(-i\omega_n + \check{\tau}_z\xi_{1\mathbf{k}}\right)\check{G}_{11} + \check{t}\check{G}_{21} = 1 \ ,\\
  \left(-i\omega_n + \check{\tau}_z\xi_{1\mathbf{k}}\right)\check{G}_{12} + \check{t}\check{G}_{22} = 0 \ ,\\
  \left(-i\omega_n + \check{\tau}_z\xi_{2\mathbf{k}}\right)\check{G}_{21} + \check{t}^{\dagger}\check{G}_{11} = 0 \ .
 \end{align}
\end{subequations}
Solving the above system, we obtain the Green's functions of the subsystems in the case of the spin-singlet interlayer pairing $\hat{\Delta}_{\rm int} = d_0(i\hat{\sigma}_y)$
\begin{subequations}
 \begin{align}
  \check{G}_{22}(\mathbf{k}) = \frac{-1}{\tilde{\omega}_{2\mathbf{k}}^2 + (\xi_{2\mathbf{k}} - \Lambda_{2\mathbf{k}})^2 + \frac{4|t|^2|d_0|^2\xi_{1\mathbf{k}}^2}{(\omega_n^2 + \xi_{1\mathbf{k}}^2)^2}}\begin{bmatrix}-i\tilde{\omega}_{2\mathbf{k}} - \xi_{2\mathbf{k}} + \Lambda_{2\mathbf{k}}&\cfrac{2t^*d_0\xi_{1\mathbf{k}}}{(\omega_n^2 + \xi_{1\mathbf{k}}^2)}(i\hat{\sigma}_y)\\ \cfrac{2td_0^*\xi_{1\mathbf{k}}}{\omega_n^2 + \xi_{1\mathbf{k}}^2}(-i\hat{\sigma}_y)&-i\tilde{\omega}_{2\mathbf{k}}+\xi_{2\mathbf{k}} -\Lambda_{2\mathbf{k}}\end{bmatrix} \ ,\\
  \tilde{\omega}_{2\mathbf{k}} = \omega_n\left(1 + \frac{|t|^2 + |d_0|^2}{\omega_n^2 + \xi_{2\mathbf{k}}^2}\right) \ , \ \ \ 
  \Lambda_{2\mathbf{k}} = \xi_{1\mathbf{k}}\left(\frac{|t|^2 - |d_0|^2}{\omega_n^2 + \xi_{1\mathbf{k}}^2}\right) \ ,\\
  \check{G}_{11}(\mathbf{k}) = \frac{-1}{\tilde{\omega}_{1\mathbf{k}}^2 + (\xi_{1\mathbf{k}} - \Lambda_{1\mathbf{k}})^2 + \frac{4|t|^2|d_0|^2\xi_{2\mathbf{k}}^2}{(\omega_n^2 + \xi_{2\mathbf{k}}^2)^2}}\begin{bmatrix}-i\tilde{\omega}_{1\mathbf{k}} - \xi_{1\mathbf{k}} + \Lambda_{1\mathbf{k}}&\cfrac{2td_0\xi_{2\mathbf{k}}}{(\omega_n^2 + \xi_{2\mathbf{k}}^2)}(i\hat{\sigma}_y)\\ \cfrac{2t^*d_0^*\xi_{2\mathbf{k}}}{(\omega_n^2 + \xi_{2\mathbf{k}}^2)}(-i\hat{\sigma}_y)&-i\tilde{\omega}_{1\mathbf{k}} + \xi_{1\mathbf{k}} - \Lambda_{1\mathbf{k}}\end{bmatrix} \ ,\\
  \tilde{\omega}_{1\mathbf{k}} = \omega_n\left(1 + \frac{|t|^2 + |d_0|^2}{\omega_n^2 + \xi_{2\mathbf{k}}^2}\right) \ , \ \ \ \Lambda_{1\mathbf{k}} = \xi_{2\mathbf{k}}\left(\frac{|t|^2-|d_0|^2}{\omega_n^2 + \xi_{2\mathbf{k}}^2}\right) \ .
 \end{align}
\end{subequations}
The poles of the resulting Green's functions together with the replacement $i\omega_n \to E$ give the quasiparticle spectrum of the two-layer system, which can be cast to the form~(8) in the main text. As a next step, we obtain the anomalous mixed Green's function $\hat{F}_{12}$, which enters the self-consistency eqauation for the interlayer order parameter (6). We get
\begin{equation}
  \hat{F}_{12} = d_0(i\hat{\sigma}_y)\frac{[-(i\omega_n + \xi_{1\mathbf{k}})(i\omega_n - \xi_{2\mathbf{k}})+|t|^2 + |d_0|^2]}{(\omega_n^2 + |t|^2 + |d_0|^2)^2 + \omega_n^2(\xi_{2\mathbf{k}}^2 + \xi_{1\mathbf{k}}^2) + \xi_{2\mathbf{k}}^2\xi_{1\mathbf{k}}^2-2\xi_{2\mathbf{k}}\xi_{1\mathbf{k}}(|t|^2 - |d_0|^2)} \ .
\end{equation}
Substituting the above expression into Eq.~(6), we obtain the gap equation
\begin{equation}
 1 = -\frac{U_0}{2}T\sum_{\omega_n}\int\frac{d^2\mathbf{k}}{(2\pi)^2}\frac{[-(i\omega_n - \xi_{1\mathbf{k}})(i\omega_n + \xi_{2\mathbf{k}}) + |t|^2 + |d_0|^2]}{[(\omega_n^2 + |t|^2 + |d_0|^2)^2 + \omega_n^2(\xi_{2\mathbf{k}}^2 + \xi_{1\mathbf{k}}^2) + \xi_{2\mathbf{k}}^2\xi_{1\mathbf{k}}^2-2\xi_{2\mathbf{k}}\xi_{1\mathbf{k}}(|t|^2 - |d_0|^2)]} \ ,
\end{equation}
which can be cast to the form~(7) in the main text via the substitution $\xi_{\mathbf{k}} = (\xi_{1\mathbf{k}}+\xi_{2\mathbf{k}})/2$.

\end{widetext}

\end{document}